\documentclass[12pt]{iopart}

\ifx\pdftexversion\undefined
  \usepackage[dvips]{graphicx}
\else
  \usepackage[pdftex]{graphicx}
\fi

\def \N{\mathcal{N}}

\begin{document}

\title[Quantum nano-electromechanics with electrons, quasiparticles and Cooper pairs]{Quantum nano-electromechanics with electrons, quasiparticles and Cooper pairs: 
effective bath descriptions and strong feedback effects}

\author{Aashish A. Clerk and Steven Bennett}  
\address{Department of Physics, McGill University,
Rutherford Physics Building, 3600 Rue University, Montr\'{e}al, QC, Canada,
H3A 2T8}

\begin{abstract}
Using a quantum noise approach, we discuss the physics of 
both normal metal and superconducting single electron transistors (SET) 
coupled to mechanical resonators. Particular attention is paid to 
the regime where transport occurs via incoherent Cooper-pair tunneling (either via the Josephson quasiparticle (JQP) or double Josephson quasiparticle (DJQP) process).  We show
that, surprisingly, the back-action of tunneling Cooper pairs (or superconducting quasiparticles) 
can be used to significantly
cool the oscillator.   We also discuss the physical origin of negative damping effects in this system,  and how they can lead to a regime of strong electro-mechanical feedback, where despite a weak SET - oscillator coupling, the motion of the oscillator strongly
effects the tunneling of the Cooper pairs.  We show that in this regime, the oscillator is characterized by an {\it energy-dependent effective temperature}.  
Finally, we discuss the strong analogy between 
back-action effects of incoherent Cooper-pair tunneling and ponderomotive effects in an optical cavity with a moveable mirror; in our case, tunneling Cooper pairs play the role of the cavity photons.  
 
\end{abstract}




\ead{aashish.clerk@mcgill.ca}

\maketitle

%\tableofcontents

%\section{Plan for paper}

%\begin{itemize}

%\item Briefly review linear response approach (emph: don't need sep of timescales between oscillator and "bath")
%\item Briefly review approach for calculating noise for JQP/DJQP
%\item Results for JQP:
%	\begin{itemize}
%		\item TEff can be really small
%		\item damping can be negative
%		\item heuristic1:  energy concerns determine sign of damping
%		\item	heuristic2:	  classical way of understanding damping (J. Harris);
%			extract effective delay time of charge response of cavity
%		\item compare damping for large and small EJ cases... different
%			delay times! (as TEff is the same)
%		\item QLimit:  emphasize no factor of g!!! much better than normal SET
%	\end{itemize}
%\item	Results for DJQP... contrast against JQP, not much discussion
%\item Discussion of unstable regime:
%	\begin{itemize}
%		\item sep. of timescales, Fokker Plank with D(x), gamma(x)
%		\item	solve... introduce TEff(E)!!!
%		\item Gaussian energy distribution (what sets width?)
%		\item predictions for zero frequency noise
%		\item	can we say anything about Sx(omega)?
%		\item noise at finite frequencies?
%	\end{itemize}
%\end{itemize}

%%%%%%%%%%%%%%%%%%%%%%%%%%%%%%%%%%%%%%%%%%%%

\section{Introduction}

There has been considerable recent interest in studying the properties of mechanical
oscillators coupled to quantum mesoscopic conductors.
Such ``quantum" nano-electromechanical systems (NEMS) 
are interesting because of their ability (in some cases)
to perform quantum-limited position detection \cite{Knobel03, LaHaye04, Mozyrsky04, Clerk04a} 
and their potential to be used in quantum control applications \cite{Ruskov05}.  In addition, they represent a new and interesting problem in
the area of quantum dissipative systems, as the tunneling electrons (or quasiparticles or Cooper-pairs) in the
conductor act as a non-equilibrium, non-gaussian dissipative bath for the 
mechanical oscillator.  Theoretically, attention has largely focused on two systems:
a normal metal single-electron transistor (SET) coupled to an oscillator
\cite{Blencowe00, Zhang02,
Chtchelkatchev04, Armour04, Blencowe04, Mozyrsky04, Blanter05, Blencowe05b}
\nocite{Blanter05b}, 
and a tunnel junction (or quantum point contact) coupled to an oscillator
\cite{Mozyrsky02, Smirnov03,Clerk04a,Ruskov05}; both systems
have also been studied experimentally \cite{Knobel03, LaHaye04, Lehnert05}.  In these devices it has
been predicted that for weak couplings, the back-action of the conductor essentially mimics the effects of an equilibrium thermal bath, with an effective temperature which is roughly proportional to the drain-source voltage in the conductor.

In this paper, we turn to the properties of a quantum NEMS system which is considerably
more complicated than either the normal-metal SET or tunnel junction systems.  We consider a superconducting single electron transistor (SSET) coupled to a nanomechanical oscillator, focusing on regimes where transport in the SSET is via the incoherent tunneling of Cooper pairs (both the Josephson quasiparticle process (JQP), and the double Josephson quasiparticle process (DJQP)) 
\cite{Averin89, Maassen91} \nocite{Averin90}.  Such processes are attractive from a measurement point of view, as they have a high gain at a relatively low source-drain voltage.  We also discuss the regime where transport through the SSET is due to the sequential tunneling of quasiparticles.
Using a quantum-noise approach,
we show that in each of these regimes the SSET provides the oscillator with a highly non-trivial effective environment.  In particular, the effective temperature of the SSET can be much lower than the drain-source voltage in the SSET, meaning that considerable cooling can be accomplished; this is in stark contrast to the normal-metal SET and tunnel junction systems, where it is difficult to achieve cooling.  In addition, the resonant nature of Cooper-pair tunneling leads to the possibility of unstable regimes characterized by negative dissipation; we discuss the strong electro-mechanical feedback that arises in these regimes and characterize the resulting stationary state of the oscillator.  This state is highly non-thermal and can be described using an energy-dependent effective temperature.  Finally, we point out that by using incoherent Cooper-pair tunneling, one can come extremely close to reaching the quantum limit on the displacement sensitivity; in contrast, this is not possible with a normal-metal SET.  Note that an SSET plus oscillator
system was also studied recently using an alternate technique by Blencowe, Imbers and Armour \cite{Blencowe05}.

The plan of this paper is as follows.  In section 2, we will review the effective bath description of a weakly-coupled NEMS, paying special attention to how 
back-action damping arises both quantum mechanically and classically.  In section 3, we introduce the SET plus oscillator system, and calculate the back-action in the sequential tunneling regime; our results here are applicable to a number of different systems, 
including a normal SET, a Coulomb-blockaded quantum dot, and a superconducting SET.  In section 4, we calculate the back-action properties of incoherent Cooper-pair tunneling, and give a physical interpretation of negative damping effects that arise here.
We point out the strong analogy between the superconducting SET system and an optical cavity with a moveable mirror, where forces arising from cavity photons
 can lead to cooling 
and instabilities \cite{Braginsky67, Metzger04, Harris05}; in our system, tunneling Cooper-pairs play the role of the photons.  Finally, in section 5, we present results for the strong feedback regime that occurs as a result of these negative damping instabilities.

%%%%%%%%%%%%%%%%%%%%%%%%%%%%%%%%%%%%%%%%%%%%

\section{Effective bath description and the origin of back-action damping}

At first glance, the theoretical 
problem presented by a NEMS system seems formidable:  we
need to understand how the back-action of our quantum conductor effects the oscillator,
given that the conductor is typically in a far-from-equilibrium state, and typically does not have gaussian noise properties.  Luckily, for many systems of interest the coupling between the oscillator and the conductor is sufficiently weak that the current in the oscillator responds linearly to the motion of the oscillator.    As is discussed extensively 
in Ref. \cite{Clerk04b}, the back-action effects of the conductor 
in this case can be directly related to the quantum noise properties of the uncoupled
detector.  We assume that the oscillator is coupled to the conductor
via a term in the Hamiltonian which is linear in $x$:
\begin{eqnarray}
	H_{int} & = & - A \hat{x} \cdot \hat{F}
	\label{eq:Hint}
\end{eqnarray}
Here, $A$ is a dimensionless coupling constant; it needs to be sufficiently weak such that the conductor only responds linearly to the motion of the oscillator.
The operator $\hat{F}$ describes the quantity in the conductor which couples to the
oscillator; 
for the SSET, we will see that $\hat{F}$ 
is simply proportional to the charge on the central island.  For simplicity,
we will choose the oscillator's origin so that $\langle \hat{F} \rangle = 0$.
$\hat{F}$ plays the role of a
fluctuating back-action force;  its unsymmetrized quantum noise spectrum at
zero coupling to the detector, 
\begin{equation}
	S_F(\omega) = \int_{-\infty}^{\infty} dt 
		e^{i \omega t} \langle \hat{F}(t) \hat{F}(0) \rangle
		,
\end{equation}
will play a central role.

Taking our oscillator to have mass $m$ and natural frequency $\Omega$, 
and assuming that it is also coupled to an equilibrium Ohmic bath,
one finds from perturbation theory in $A$ that its motion is described by
a classical Langevin equation \cite{Clerk04b}:
\begin{eqnarray}
    m \ddot{x}(t)  =  -m \Omega^2 x(t) - m \gamma_0 \dot{x} + \delta f_0(t) 
     + f_{avg}(t) +
    \delta f(t)
    \label{eq:Langevin}
\end{eqnarray}
This classical Langevin equation involves two fluctuating forces, $\delta f_0(t)$ and $\delta f(t)$; it may be 
used to calculate the fluctuations in $x(t)$ in terms of the spectral densities of the $\delta f_0$ and $\delta f$ 
fluctuations, as well as the oscillator's response to an external force.  {\it The answers 
obtained for both these quantities from Eq.~ (\ref{eq:Langevin}) are in exact correspondence to 
what is found from a perturbative quantum mechanical calculation, 
performed to lowest non-vanishing order in $A$} \cite{Clerk04b}.

Consider the RHS of this Langevin equation.  The second and third terms on the RHS describe the effects of the equilibrium bath on the oscillator:  $\gamma_0$
is the damping rate induced by this bath, and $\delta f_0(t)$ is the corresponding fluctuating force.  The fluctuation-dissipation theorem demands that the spectral density of the $\delta f_0$ fluctuations obey:
\begin{equation}
	S_{\delta f_0}(\omega) = m \gamma_0 \hbar \omega \coth
		\left( \hbar \omega / (2 k_B T_0) \right)
\end{equation}
where $T_0$ is the bath temperature.  $\gamma_0$ and $T_0$ have a simple interpretation:  if the oscillator were not coupled to the conductor, its quality factor would be $Q_0 = \Omega / \gamma_0$ (the ``intrinsic" quality factor of the oscillator) and its temperature would simply be $T_0$.

The remaining terms on the RHS of Eq.~ (\ref{eq:Langevin}) describe the back-action of the conductor.
$\delta f(t)$ is the fluctuating part of the back-action force; its spectral density is determined
directly by the symmetrized quantum noise in the operator $\hat{F}$:
\begin{eqnarray}
	S_{\delta f}(\omega) = \frac{A^2}{2}\left(
		S_F(\omega) + S_F(-\omega) \right)
		\label{eq:FClassSpectrum}
\end{eqnarray}
In contrast, $f_{avg}(t)$ is the average value of the back-action force; it arises
because the conductor (and hence $\langle \hat{F} \rangle$) changes in response to the motion of the oscillator.  We
can write this force as the sum of a conservative force which is 
in-phase with the oscillator's motion, and a damping force which is out-of-phase with the oscillator's motion:  
\numparts
\begin{eqnarray}
	f_{avg}(t) & = & f_{damp}(t) + f_{cons}(t) \\
	f_{damp}(t) & = & 
		-m\int_{-\infty}^{\infty} dt'\gamma(t-t')\dot{x}(t')
		\label{eq:Fdamping}\\
	f_{cons}(t) & = & 
		\int_{-\infty}^{\infty} dt' \alpha(t-t'){x}(t')
		\label{eq:Fconservative}
\end{eqnarray}
\endnumparts
Using standard quantum linear response relations, one has:
\numparts
\begin{eqnarray}
 	\lambda_{F}(t) & = & -\frac{i}{\hbar}\theta(t)\langle[F(t),F(0)]\rangle. \\
 	m\gamma(\omega) & = & A^{2}\left(
		\frac{- \textrm{Im} \lambda_{F}(\omega)}
			{\omega}\right)
			=
			\frac{A^{2}}{\hbar}\left(\frac{S_{F}(\omega)-S_{F}(-\omega)}{2\omega}\right)
	\label{eq:DampingKernel} \\
	\alpha(\omega) & = & A^{2}\left(- \textrm{Re} \lambda_{F}(\omega)\right)
	\label{eq:AlphaKernel} 
\end{eqnarray}
\endnumparts
Note that this description of the oscillator only requires $A$ to be small enough that linear response is valid; 
there is no restriction on the size of the oscillator frequency $\Omega$.  Note also
that the fluctuating back-action force $\delta f(t)$ will not in general be related to the
damping kernel $\gamma(\omega)$ by a fluctuation-dissipation
relation, as our conductor is generally not in an equilibrium state.  Nonetheless, we may use the fluctuation dissipation relation to {\it define} an effective temperature $T_{eff}(\omega)$ at each frequency $\omega$ via:
\begin{equation}
	\coth
		\left(
		\frac{ \hbar \omega}{ 2 k_B T_{eff}(\omega) } \right) 
		\equiv
		\frac{S_{\delta f}(\omega) }{m \gamma \hbar \omega}
		\label{eq:FullTEff} 
\end{equation}
In the most general case, the oscillator responds to a variety of frequencies, and thus we cannot characterize the
back-action by a single temperature; in this case, the oscillator will not be in a thermal state.  However, if the oscillator has a sufficiently high quality factor, it only responds to forces in a narrow frequency band centered
on $\omega = \Omega$.  If this frequency band is small enough, $T_{eff}(\omega)$ will be constant over its width.
In this case, the back-action will appear to the oscillator as being due to an equilibrium bath having
a temperature $T = T_{eff}(\Omega)$ and a damping rate $\gamma(\Omega)$.  The spring constant $k=m \Omega^2$ of the oscillator will also be modified by the conservative part
of the average back-action force; this will be given by:
\begin{equation}
	\Delta k = - \alpha(\Omega) = A^2 (\textrm{Re } \lambda_F(\Omega) )
\end{equation}
Including the effects of the equilibrium bath coupled to the oscillator, Eq.~ (\ref{eq:Langevin}) tells us that the oscillator will be in a thermal state,
characterized by a total damping rate $\gamma(\Omega) + \gamma_0$ and a temperature $T_{osc}$ given rigorously by \cite{Clerk04b}:
\begin{eqnarray}
 	T_{osc} &= & 
		\frac{\gamma_{0}T_{0}+\gamma(\Omega) 
			T_{eff}}{\gamma_{0}+\gamma(\Omega)}
		\label{eq:TOsc}
\end{eqnarray}

We again stress that the effective bath description presented above can hold even if the oscillator
frequency $\Omega$ is not small 
compared to the frequency scales of the conductor; all that is needed is that the 	``mechanical bandwidth" of the oscillator be small.
Nonetheless, we are often interested in the case where $1/\Omega$ is long compared to the
timescales relevant to the conductor; in this case, we can approximate 
$\gamma(\omega) \simeq \gamma(0)$, $\alpha(\omega) \simeq \alpha(0)$ and $S_{\delta f}(\omega) \simeq S_{\delta f}(0)$.  This corresponds to a damping force $F_{damp} = -m \gamma(\omega=0) \dot{x}$, a modified spring constant $\Delta k = - \alpha(\omega=0)$, and a fluctuating back-action force with a white
noise spectrum.  In this case, Eq.~ (\ref{eq:FullTEff}) reduces to:
\begin{eqnarray}
 	k_{B}T_{eff} \equiv \frac{S_{F}(0)}{2\partial_{\omega}S_{F}(0)/\hbar}=\frac{A^{2}S_{F}(0)}{2m\gamma}\label{eq:Teff}\end{eqnarray}

Before applying these ideas to the SET plus oscillator system, we comment on the origin of back-action damping, as there seems to have been some confusion on this point in the NEMS literature. 
Eq.~ (\ref{eq:DampingKernel}) makes it clear that there are two equivalent ways of thinking about damping. 
The first equality tells us that damping is simply the out-of-phase response of the average
value of the back-action force $F$ to the position of the oscillator. The second equality gives us a quantum
picture for damping:  damping results from the absorption of energy from the oscillator
by the {}``bath'' producing the force $F$ (recall that positive-frequency
noise corresponds to absorption of energy, while negative frequency
noise corresponds to emission of energy \cite{Schoelkopf03}).

There is another simple but useful classical way to understand both the origin of back-action damping and the spring-constant renormalization \cite{Braginsky67, Metzger04, Harris05}.  We simply need to use the fact that a) the average back-action force $f_{avg}(t)$ is proportional to $x$, and b) the back-action force does not respond instantaneously to changes in the oscillator's motion.  In the simple case where the oscillator is much slower than the source
of the back-action, we can take the time delay to be given exactly by $\tau_{resp}$
	\footnote{
		A more exact equation would be to say $f_{avg}(t) = 
		\frac{df_{avg}}{dx} \int_{-\infty}^{t} dt'
		\exp[-(t-t')/\tau_{resp}] x(t')$.  For $\Omega \tau_{resp} \ll 1$, this reduces to Eq.~ 
		 (\ref{eq:ClassDamping}) above.}.
We thus have:
\begin{eqnarray}
	f_{avg}(t) & = & f_{avg}[x(t-\tau_{resp})] \nonumber \\
		& \simeq &	\frac{d f_{avg}}{dx} \cdot x(t-\tau_{resp}) \nonumber \\
		& \simeq &	\frac{d f_{avg}}{dx} \cdot \left[ x(t)-\tau_{resp} \dot{x}(t)\right] 
		\label{eq:ClassDamping}
\end{eqnarray}
This immediately yields:
\numparts
\begin{eqnarray}
	\Delta k = - \frac{df_{avg}}{dx} \\
	m \gamma = \frac{df_{avg}}{dx} \tau_{resp}
\end{eqnarray}
\endnumparts
Thus, we can view the time delay in the response of the back-action force as being the origin of damping.  Note that this argument tells us that $\Delta k$ and $\gamma$ should always have the opposite sign, and that their ratio gives us a measure of $\tau_{resp}$, the response time of the conductor's back-action:
\begin{equation}
	\tau_{resp} \equiv  - \frac{m \gamma}{\Delta k}
	\label{eq:taudefn}
\end{equation}
As we will see, the concept of a back-action response time will be useful in understanding the
back-action properties of various different mesoscopic systems.

%Using these relations, we may now calculate the back-action of an
%SSET on an oscillator by simply calculating the noise properties of
%the oscillator. A convenient method for doing this is discussed in
%Ref. \cite{ref:QubitsSpectrometers}, where one studies a coupled
%system of a qubit plus SSET, and uses the long-time behaviour of the
%qubit to determine the symmetric and asymmetric-in-frequency parts
%of the SSET quantum noise. In what follows, we discuss results obtained
%from this approach for both the JQP and DJQP resonances, highlighting
%important differences between the two cases. We also try to provide
%intuitive explanations for the structure of the results. Note that
%the back-action properties of the DJQP process, in the context of
%qubit state detection, were previously studied in Ref. \cite{ref:DJQPPRL}.

%%%%%%%%%%%%%%%%%%%%%%%%%%%%%%%%%%%%%%%%%%%%

\section{Back-action of electron and quasiparticle sequential tunneling}

A SET consists of a metallic island with a large Coulomb charging
energy 
$E_C = e^2 / (2 C_{\Sigma})$ ($C_{\Sigma}$ is the total capacitance of the island) 
coupled via tunnel junctions to both a source and a drain 
metal electrode; in the case of a SSET, the island and both the electrodes are superconducting.  The charging-energy term in the Hamiltonian is given by:
\begin{eqnarray}
	H_C = E_C(\hat{n} - \mathcal{N})^2,
	\label{eq:HC}
\end{eqnarray}
where $\hat{n}$  is the charge on the SET island, and
$\mathcal{N} = C_g V_g / e$ is the dimensionless electron number associated
with a gate voltage $V_g$ which is coupled to the island via a capacitance $C_g$.
In addition, a voltage $V_{SD}$ is applied between source and drain which drives
the tunneling of electrons across the SET.

To make a NEMS device involving a SET, a mechanical oscillator (coated with metal) is placed in proximity to the island of the SET such that there is a capacitive coupling $C_{osc}$ between the charge on the SET island and the potential of the oscillator;
$C_{osc}$  depends on the island-oscillator distance.  The oscillator is then voltage biased, with the result that changes in its displacement modify the 
electrostatic potential of the SET island (i.e. the parameter $\mathcal{N}$ becomes $x$ dependent).  For small displacements $x$, the result is a linear-in-$x$ coupling term in the Hamiltonian
	\footnote{Note there is an additional term in the SET plus oscillator
	 hamiltonian which is 	proportional to $V_{osc}^2 x^2$, and 
	 which thus contributes a $V_{osc}$-dependent
	shift in the oscillator's frequency.  This term has nothing to do with back-action:  it
	is independent of the SET island charge $\hat{n}$.  As such, we do not discuss it 
	in what follows.}
:
\begin{eqnarray}
 H_{int} & = & -A\hat{n}x
 	\label{eq:HintSET}
 \end{eqnarray}
where $A$ is the effective coupling strength:
\begin{eqnarray}
	A  =  	
		2 E_C \frac{ d \mathcal{N} }{d x} & = & 
		2E_{C} \left(\frac{V_{osc}} {e} \right) 
			\frac{dC_{osc}}{d x} 
	\nonumber \\
		& \simeq & 
			2E_{C} \left(\frac{ C_{osc} V_{osc}} {e}\right)    
			 \frac{1}{d} 
	\nonumber \\
		& = & 
			\e V_{osc} \left(\frac{ C_{osc} } {C_{\Sigma}}\right)    
			 \frac{1}{d}. 
			\label{eq:SETA}
\end{eqnarray}
Here, $V_{osc}$ is the voltage applied to the oscillator, and $d$ is the 
oscillator-island spacing when the oscillator is in its equilibrium position; for
more details, see, e.g.,  Ref. \cite{Blencowe04}.
Comparing against Eq.~ \ref{eq:Hint}, we see that $A \hat{n}$ plays the role
of a fluctuating back-action force on the oscillator, and that it is the quantum noise spectrum of the charge $\hat{n}$ that will determine the nature of this back-action.  This conclusion
is true regardless of where the SET is operated, or whether it is
superconducting or normal.  

Calculating the back-action thus reduces to a problem of calculating 
the quantum noise spectrum of charge fluctuations in the SET.  For various operating points of a SET/SSET, this has been done 
\cite{Korotkov94, Averin00, Johansson02, Maassen02, Clerk02, Schoelkopf03}.  The simplest operating regime of a SET is that of sequential tunneling.  
Current flow is due to electrons (or, in the case
of a SSET, quasiparticles) incoherently tunneling on and off the SET island via
real, energy conserving transitions; because of the large charging energy $E_C$, only two charge states of the island are involved in transport.
The quantum charge noise in this regime was calculated in \cite{Schoelkopf03} by 
using a qubit as a theoretical quantum noise spectrometer.  In this approach,  one
first models the dynamics of a two-level system (TLS) weakly coupled to the SET island charge, and calculates an equation of motion for a reduced density matrix tracking both the TLS state and the SET island charge;
for sequential tunneling, this is done to lowest non-vanishing order in the tunneling in the SET.
In the limit of weak SET-TLS coupling, one can 
calculate both the stationary state of the TLS, as well as its relaxation rate.  These may then be used to directly extract the finite frequency quantum noise spectrum of $\hat{n}$
at the splitting frequency of the TLS.  By varying this splitting frequency, one
can extract the entire quantum noise spectrum of $\hat{n}$.  

Using the above approach, one finds 
that the low-frequency SET charge noise in the sequential tunneling regime is simply telegraph noise, while the frequency-asymmetry in the noise is determined by how the tunneling rates depend on energy \cite{Johansson02, Clerk02}.  To be definite, consider
a sequential tunneling process involving the island charge states $n=0$ and $n=1$, and
consider the zero temperature limit.  There are then two rates of interest: $\Gamma_{+}$, the rate at which electrons (or quasiparticles) hop onto the island from the left junction, and $\Gamma_{-}$, the rate at which electrons hop off
the island through the right junction.  The average current is given by 
$I = e \Gamma_{+} \Gamma_{-}/\Gamma_{\Sigma}$ 
(where $\Gamma_{\Sigma} = \Gamma_+
+ \Gamma_-$), and the average island is $\langle n \rangle = e \Gamma_+ / \Gamma_{\Sigma}$.  It will also be important to know how these rates
change if we either increase or reduce the amount of energy driving the given tunnel event; let $E$ represent this additional added energy.
Using Eqs. (\ref{eq:DampingKernel}) and (\ref{eq:Teff}) and the results of 
Ref. \cite{Schoelkopf03},  one obtains the  following simple expressions for the back-action in the limit of small oscillator frequency (i.e. $\Omega \ll \Gamma_{\Sigma}$):  
\numparts
\begin{eqnarray}
	k_B T_{eff} & = & 
		\frac{\Gamma_{+} \Gamma_{-}} 
		{ \partial_E \left( \Gamma_{+} \Gamma_{-} \right)} \\
	m \gamma & = &
		A^2 \frac{
			\partial_E \left( \Gamma_{+} \Gamma_{-} \right)}
			{ \left(\Gamma_{\Sigma}\right)^3 }  \\
	\delta k	& = & -(m \gamma) \times \Gamma_{\Sigma}
\end{eqnarray}
\endnumparts
Here, each rate should be evaluated at $E=0$.  These 
expressions are equally valid for sequential tunneling of electrons in a SET,
electrons in a Coulomb-blockaded quantum dot, or BCS quasiparticles in a 
superconducting SET
	\footnote{
		Note that there is a difference between a normal-metal SET and a Coulomb
		blockaded quantum dot:  the former system has a vanishing single-particle
		level spacing, while the latter system has a large single-particle level spacing.
		As a result, the tunneling rates and back-action are quite different.};
the only difference is in the form of the tunnel rates.  Note that the bath response time $\tau_{resp}$ defined in Eq.~ (\ref{eq:taudefn}) is simply given by 
$\tau_{resp} = 1/(\Gamma_{\Sigma})$, as could have been expected:  this is the timescale which characterizes charge relaxation on the SET island.

For a normal-metal SET, in the case where both junctions
have a conductance $g e^2 / h$ and equal capacitances, 
the tunnel rates are given by
\begin{eqnarray}
 	\Gamma_{\pm}(E) = 
		\frac{g (E_{\pm} + E)}{h}
		 \left[ 1 + n_B(E_{\pm}+E) \right],
\end{eqnarray}
where $n_B$ is the Bose-Einstein distribution function, and $E_{\pm}$ is the gain of electrostatic energy associated with each tunneling process:
\numparts 
\begin{eqnarray}
	E_+ & = & 
			e V_{ds}/2 - 2 E_C (1/2 - \mathcal{N})  \\
	E_- & = & 
			e V_{ds}/2 + 2 E_C (1/2 - \mathcal{N}) 
\end{eqnarray}
\endnumparts
As a result, at zero physical temperature in the SET, the effective temperature and back-action damping are given by:
\numparts
\begin{eqnarray}
	k_B T_{eff} & = & 
		p_0 p_1 eV_{ds} \\
	m \gamma & = & A^2 \cdot  \frac{h}{g (eV_{ds})^2} 
\end{eqnarray}
\endnumparts
where $p_0 = \Gamma_{-} / \Gamma_{\Sigma}$ is the probability of having $n=0$, $p_1 = 1-p_0$ is the probability of having $n=1$.  For a fixed $V_{ds}$, the damping is constant,
while the effective temperature reaches its maximum $e V_{ds}/ 4 k_B$ at the point of
maximum current, $\mathcal{N} = 1/2$.  Note these results are in agreement with Refs. \cite{Armour04, Blencowe05b}, which treats the SET-oscillator system using a generalized
master equation.

Turning to the case of a superconducting SET (i.e. SSET) with identical junctions, the quasiparticle tunneling rates are now given by:
\begin{eqnarray}
	\Gamma_{\pm} & = & 
		I_{SS}
		\left[ (E_{\pm}+E) / e \right]
		\left[ 1 + n_B(E_{\pm}+E) \right]
	\label{eq:QPRate}
\end{eqnarray}
Here,  $I_{SS}(V)$ is the usual I-V characteristic of a SIS junction \cite{Tinkham96}:  there is no current until $V= 2\Delta/e$, after which there is an abrupt rise.  In the usual case where there is
some inelastic scattering in the superconductor, this discontinuity in the I-V curve
 is smeared out
over a voltage $h/ (e \cdot \tau_{QP})  \ll \Delta/e$, where $\tau_{QP}$ is the quasiparticle lifetime in the superconductor \cite{Dynes78}.
	\footnote{Note that environmental voltage fluctuations in the SET will have a similar
	effect of smearing out the energy dependence of the
	quasiparticle tunnel rates, and can also be parameterized by $\tau_	{QP}$.}
Note that the threshold voltage for quasiparticle
sequential tunneling is $V_{DS} = 4 \Delta$, as one needs to create four quasiparticles
to transfer a charge $e$ from left to right.  To compare against the normal SET case, we further specialize to the case where
$\mathcal{N} = 1/2$; this gives a maximum current.  
We now have for both normal and superconducting SET's that $\Gamma_{+} = 
\Gamma_{-} \equiv \Gamma$, and:
\numparts
\begin{eqnarray}
	k_B T_{eff} & = & 
			\frac{ \Gamma }
		{ 2 \cdot \partial_E \Gamma} \\
	m \gamma & = & 
		A^2 \frac { \partial_E \Gamma}{4  \Gamma^2}
\end{eqnarray}
\endnumparts 
For drain source voltages much larger than $4 \Delta / e$, the tunnel rates for the quasiparticles will be almost identical to that for normal electrons, and the back-action 
will be almost identical to that in a normal SET at a similar voltage.  However, for $e V_{DS} \sim 4 \Delta$, the sharp rise of the SIS current-voltage characteristic will lead to
stronger back-action in the superconducting case.  One finds that 
$\partial_E \Gamma$ will be larger than in the normal state case by a large
factor $\Delta \tau_{QP} / h$.  One thus has:
\numparts
\begin{eqnarray}
	k_B T_{eff} & \sim & \frac{h}{\tau_{QP}} \\
	m \gamma & \sim & 
		A^2 \frac { \tau_{QP}}{\Delta  }
\end{eqnarray}
\endnumparts 
Thus, near threshold, quasiparticle sequential tunneling can lead to very low effective temperatures and high damping rates.  This behaviour is entirely due to the sharp dependence of the quasiparticle tunneling rates on energy, and is limited by this
sharpness.  Note that unlike a normal SET or a tunnel junction NEMS, the scale
of $T_{eff}$ here is not simply set by the drain-source voltage.

%%%%%%%%%%%%%%%%%%%%%%%%%%%%%%%%%%%%%%%%%%%%

\section{Back-action of incoherent Cooper-Pair tunneling}

\begin{figure}
\center{\includegraphics[width=14 cm]{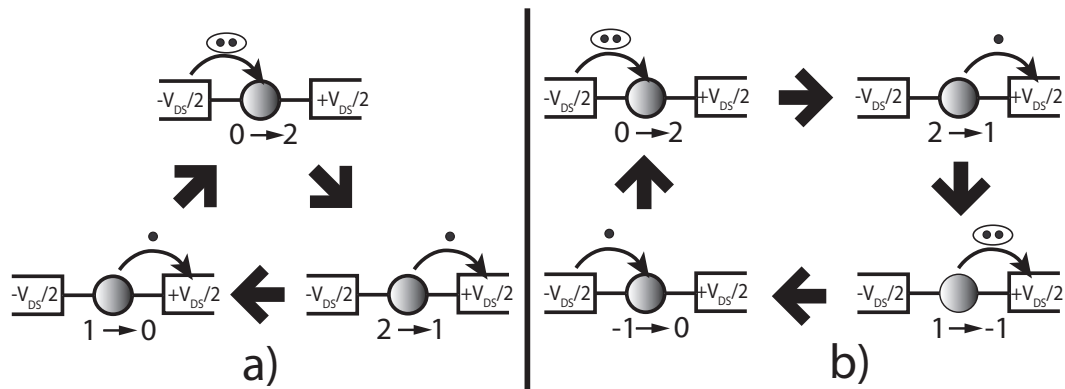}}
\vspace{-0.5cm}
\caption{\label{fig:DJQP}
a) Schematic of the JQP process, showing each of the three charge transfer
steps.  b) Schematic of the DJQP process.
}
\end{figure}

\subsection{JQP Process}

We now examine the back-action of incoherent Cooper-pair tunneling processes
in a SSET.  The simplest such process is the Josephson quasiparticle resonance (JQP), a transport cycle which has been studied extensively in the context of superconducting
qubits \cite{Averin89,Maassen91,Nakamura96,Choi01}\nocite{Choi03}.  The first step of a JQP cycle involves a Cooper-pair tunneling 
through one junction of the SET,
changing the charge state of the SET island by two electrons.  This is followed by two quasiparticle tunneling events through the other junction which return the SET island to its initial charge state; the cycle then repeats (see Fig.~\ref{fig:DJQP}a) .  Unlike sequential tunneling (of either electrons or quasiparticles), incoherent Cooper-pair tunneling is a quantum resonant process; as such, its back-action will have unique signatures.

For definiteness, we focus on a JQP resonance where 
a Cooper pair first tunnels onto the island from the left, taking the
island charge from $n=0$ to $n=2$. We then have a quasi-particle
tunneling event involving the right lead, taking the island charge
from $n=2$ to $n=1$ (described by a rate $\Gamma_{a}$). Finally,
another quasiparticle tunneling occurs in the right lead, taking $n=1$
to $n=0$ (described by a rate $\Gamma_{b}$). There are three important
energy scales here: $ $$\delta,$ the detuning of the Cooper-pair
resonance, and $E_{a}$($E_{b})$, the amount of energy driving the
quasi-particle event $\Gamma_{a}$($\Gamma_{b})$. In the case of symmetric
SSET junctions, these energies
are given simply by:
\numparts
\begin{eqnarray}
\delta & = & 
	E_{final} - E_{initial} = 
	4E_{C}(1-\mathcal{N})-eV_{DS}\label{eq:DeltaDefn}\\
E_{a} & = & 2E_{C}(3/2-\mathcal{N})+eV_{DS}/2
	= \delta/2 + eV_{DS} + E_C
\label{eq:EaDefn}\\
E_{b} & = & 2E_{C}(1/2-\mathcal{N})+eV_{DS}/2
	= \delta/2 + eV_{DS} - E_C
	\label{EbDefn}
\end{eqnarray}
\endnumparts
The modification for non-symmetric junctions is straightforward, 
see e.g. \cite{Esteve92}.
The quasiparticle tunneling rates $\Gamma_{a},$$\Gamma_{b}$
are determined from $E_a$, $E_b$ and Eq.~ (\ref{eq:QPRate}).  Note that as $E_a > E_b$, one always has $\Gamma_a > \Gamma_b$.

To calculate the current near the JQP feature, the standard approach is
to calculate the equation of motion for a reduced density matrix describing the 
SSET island charge; this is done to lowest non-vanishing order in perturbation
theory in tunneling in the junctions \cite{Averin89}.  The approximation here is that the tunneling is weak, which requires the dimensionless conductance $g$ of the junctions to be much smaller than $2 \pi$ (we consider the case where both junctions have equal
conductances for simplicity).  This density-matrix approach can be extended
to calculate the quantum charge noise of the island, as discussed in Ref. \cite{Clerk02}.
One again uses the ``qubits as spectrometer" idea discussed in the previous section, 
and studies a system where the SSET island is coupled to a two level system (TLS).  In complete analogy
to the calculation for the charge noise of sequential tunneling \cite{Schoelkopf03},  the weak-coupling, long time dynamics of this system, evaluated for different TLS splitting frequencies, 
yields the quantum noise spectrum of $\hat{n}$.

In what follows, we present results for the transport and noise properties for the JQP cycle obtained from this approach.  Three main interesting effects are found:  i) JQP can lead to an extremely low effective temperature, meaning that it can be used for active cooling; ii) JQP can lead to negative damping effects; the resulting instability leads to an interesting regime of strong electro-mechanical feedback (to be described in detail in Sec. 5); iii) JQP can be used to reach the quantum-limit of position detection within a factor of two.

The current for the JQP process is given by \cite{Averin89}:
\begin{equation}
	I[ \mathcal{N}, V_{ds}] = 	2e
		\frac{E_J^2 \Gamma_a }
			{4 \delta^2 + (\hbar \Gamma_a)^2 + E_J^2(2+\Gamma_a / \Gamma_b)}
	\label{eq:JQPFullAvgI}
\end{equation}
Here, $E_J = g \Delta / 8$ is the Josephson energy which sets the strength of coherent Cooper-pair tunneling.  Note that both  $\Gamma_{a}$ and $\Gamma_{b}$ in the above equation depend only weakly on $\mathcal{N}$ and $V_{DS}$ near the JQP resonance; the main dependence of the current on these parameters is through 
$\delta$, the energy detuning of the Cooper-pair resonance.  To obtain a heuristic understanding of the JQP cycle, it is useful to consider
Eq.~ (\ref{eq:JQPFullAvgI}) in the limit of small $E_J / (\hbar \Gamma_a)$; this limit is often approached in real experiments.  In this limit, Cooper-pair tunneling is the limiting step of the JQP cycle, and one finds the simple result that $I$ is $2e$ times the incoherent Cooper-pair tunneling rate from the charge state $n=0$ to $n=2$:
\begin{eqnarray}
	I & = & 2e \times \Gamma_{CPT}(\delta) = 
		2e \times
		\left( \frac{E_J}{2} \right)^2 
		\frac{\Gamma_{a}}
			{\delta^{2}+(\hbar \Gamma_{a}/2)^{2}}
	\label{eq:JQPAvgI}
\end{eqnarray}
The incoherent Cooper-pair tunneling rate has the usual form expected from Fermi's Golden rule: a transition matrix element squared (i.e. $(E_J/2)^2$) times a density of states.  Here the density of states is that of the final $n=2$ charge state, which is lifetime-broadened by the quasiparticle transition $\Gamma_a$.  

%It also useful to consider the opposite limit, where Cooper-pair tunneling is much faster than quasiparticle tunneling.  For $E_J \gg \Gamma_a \gg \Gamma_b$, we obtain the simple expression:
%\begin{equation}
%	I = 	2e \Gamma_b \cdot
%		\frac{\left(E_J/2\right)^2  }
%			{\delta^2 +  \left(E_J/2\right)^2}
%	\label{eq:JQPFullAvgI}
%\end{equation}
%In this limit, the island charge is not a good quantum number; rather, one should work with the eigenstates of the system at zero quasiparticle tunneling (i.e. superpositions of the $n=0$ and $n=2$ charge states).  As $\Gamma_b \ll \Gamma_a$, we can think of quasiparticle tunneling as a single incoherent process taking the $n=2$ charge state to $n=0$.  This process will lead to transitions which start and end with the same given eigenstate [WHY?].  The current is equal to 2e times the rate of such a transition; the latter is proportional to both the QP transition rate ($\Gamma_b$) and the appropriate matrix element (second factor above).

We turn now to the back-action of the JQP process.  For simplicity, we will focus on
operating points near the center of the JQP resonance, and make use of the smallness of the dimensionless conductance $g$.  In this regime, one can safely neglect the quasiparticle contributions to back-action damping (i.e. $\partial_E \Gamma_a$, $\partial_E \Gamma_b$), as they are higher order in $g$.  The dominant contribution to back-action damping will instead arise from the oscillator's ability to modify the condition for resonance; as we will discuss, this is in complete analogy to damping in a number of different quantum resonant systems.  The neglect of quasiparticle damping will allow us to derive some simple analytic expressions characterizing the back-action.

In the usual limit of a small oscillator frequency (i.e. $\Omega \ll \Gamma_a, \Gamma_b, E_J$), the back-action of an SSET biased near the JQP resonance is described by:
\numparts
\begin{eqnarray} 	
	k_{B}T_{eff} & = &
			\frac{(\hbar \Gamma_{a})^{2}+4\delta^{2}}{16\delta}
%		=
%			\frac{1}{2\partial_{\delta}\log\left(\gamma_{CPT}(\delta)\right)}
		\label{eq:JQPTeff}	\\
	\frac{m \gamma}{A^2} & = & 
			16 \delta \cdot  \frac{E_J^2 \Gamma_a}{\Gamma_b^2} 
				\left[ \frac{
						4 \delta^2 +  c_1 (\hbar \Gamma_a)^2 + c_2 E_J^2 }
					{\left(
						4 \delta^2 + (\hbar \Gamma_a)^2 + c_3 E_J^2 \right)^3 }
				\right] 	
			\label{eq:JQPgamma} \\
	\tau_{resp}  & \equiv&  
			\frac{- m \gamma}{\Delta k} =
			\frac{\Gamma_a}
				{\Gamma_{b} \left(\Gamma_a + 2 \Gamma_b\right)} 
					\cdot
				\frac
				{4 \delta^2 + c_1 (\hbar \Gamma_a)^2 +  c_2 E_J^2}
				{4 \delta^2 + (\hbar \Gamma_a)^2 + c_3 E_J^2} 
		\label{eq:JQPtau}
\end{eqnarray}
\endnumparts
where the dimensionless coefficients $c_1, c_2$ and $c_3$ are determined by the asymmetry between the two quasiparticle tunnel rates:
\numparts
\begin{eqnarray}
	c_1  & = & 1 + 4 \frac{\Gamma_b}{\Gamma_a} +
		8 \left(\frac{\Gamma_b}{\Gamma_a}\right)^2 \\
	c_2 & = & 1 + 4 \frac{\Gamma_b}{\Gamma_a} +
		4 \left(\frac{\Gamma_b}{\Gamma_a}\right)^2 \\
	c_3 & = & 2 + \frac{\Gamma_a}{\Gamma_b}
\end{eqnarray}
\endnumparts
Note that the above expressions are valid for {\it arbitrary} ratios of $E_J / \Gamma_a$ and $\Gamma_a / \Gamma_b$.  

Several comments are in order.  First, note that whenever the detuning of the Cooper-pair resonance is negative (i.e. $\delta < 0$), both $T_{eff}$ and $\delta$ become negative.  This behaviour has a simple physical interpretation \cite{Clerk02}.  
From Eq.~ (\ref{eq:DeltaDefn}), we see that $\delta>0$ means
that $V_{DS}$ is \emph{smaller} than what is needed to have the Cooper-pairs
on resonance.  In this case, if the tunneling Cooper-pairs can absorb energy from the oscillator they can move closer to resonance, while if they emit energy to the oscillator, they will move further from resonance.  The net result is that the  SSET prefers to \emph{absorb} energy from the oscillator, which implies both $\gamma >0$, $T_{eff} > 0$; this is the same behaviour exhibited by an equilibrium bath.  In contrast,
when $\delta < 0$, the situation is reversed.  $V_{DS}$ is now larger than what is needed to have the Cooper pairs on resonance, and the SSET prefers 
to \emph{emit} energy to the oscillator in order to move the Cooper pairs closer to resonance.  This is the meaning of negative damping and negative temperature:
our effective bath (i.e. the SSET) prefers to $excite$ the oscillator rather than to absorb
energy from it.  Using Eq.~ (\ref{eq:TOsc}), we see that the effect of negative damping is to {\it reduce} the total damping
compared to that provided by the equilibrium bath, and {\it increase} the oscillator temperature.  For
$\gamma + \gamma_0 < 0$, the total damping of the oscillator becomes negative, and we enter an interesting unstable regime.  The properties of this regime are the subject of Sec. 5.  

%Several comments are now in order.  First, note that $T_{eff}$ is set by the 
%detuning $\delta$ of the Cooper-pair resoance and the quasiparticle tunneling rate $\Gamma_{a}$ (i.e. the inverse lifetime of the $n=2$ charge state).  The minimum $|T_{SSET}|$ of $\Gamma_{a}/4$ is obtained for a detuning $|\delta| = \Gamma_a/2$; at this point, the current the JQP is still large (i.e. at most a factor of 2 small than its maximal value).  As the dimensionless conductance of the junctions could be made quite small, this minimum temperature can be much smaller than $V_{ds}$.  This behaviour is marked contrast to results for a normal SET in the sequential tunneling regime (c.f. Eq.~ XX), or for tunnel junction / quantum point contact.  In both those systems,  $T_{eff}$ is on the order of the drain source voltage $V_{ds}$ (unless the current is made small).  Thus, the idea that the effective temperature of a current-carrying mesoscopic conductor is always set by the voltage is patently false!
%
%The extremely simple expression for $T_{eff}$ is a consequence of the resonant nature of transport in the JQP cycle.  A completely analogous expression for the effective temperature
%is found for other resonant non-equilibrium baths.  A particularly simple example is the pondermotive cooling of a moveable mirrror in a resonant cavity by photons in the cavity; 
%in this case $T_{eff}$ is again given by Eq.~ XX, with $\Gamma$ being replaced by $1/\tau$, where $\tau$ is the lifetime of photons in the cavity [CITE].  
%
The fact that the sign of $\gamma$ and $T_{eff}$ change as we tune the detuning $\delta$ 
through the resonance is a generic feature of the noise properties of a number of different resonant
systems.  For example, identical effects occur in an optical Fabrey-Perot cavity 
where one mirror is flexible \cite{Braginsky67, Metzger04, Harris05}.  If such a cavity is driven by laser light which is slightly detuned from the frequency of the cavity, the resulting ``bath" of cavity photons will both damp and heat the mirror.  The detuning parameter in this case is $\delta = \hbar( \omega_{cavity} - \omega_{laser})$; when $\delta < 0$, photons from the laser can become resonant with the cavity {\it if} they give up energy to the mirror, and one gets negative damping.  
%Similar effects also occur in the Doppler cooling of atoms [CITE]; in that case, changing %the detuning of the laser with respect to the atomic transition frequency also changes %the sign of the effective damping.
This analogy can in fact be taken further:  {\it the effective temperature for the optical cavity system is identical to the expression found for JQP}, c.f. Eq.~ (\ref{eq:JQPTeff})
\cite{Harris05}.
For the JQP process, $T_{eff}$ is set by the detuning $\delta$ and the quasiparticle rate $\Gamma_a$; $\Gamma_a$ is the inverse lifetime of the resonant $n=2$ charge state.  The expression for the optical cavity system is identical, with $\Gamma_a$ being replaced by the inverse lifetime of a photon in the cavity \cite{Harris05}.
%\footnote{The expression for $T_{Eff}$ in the Doppler cooling case is in fact twice 
%as large as Eq.~ (\ref{eq:JQPTeff}),
%as one must also include the effects of transitions where the atom emits a photon}.  
The correspondence is a direct consequence of the fact that in both cases, the motion of a mechanical degree of freedom modifies a 
resonance condition.  

While the analogy to the optical cavity system is certainly useful, there are some respects in which it is different from JQP case.  Consider the expression for the response time $\tau_{resp}$ of our effective bath.  As expected, $\tau_{resp}$ scales as the lifetime of the resonant state (i.e. $1/\Gamma$).  More surprising is the fact that $\tau_{resp}$ depends on the detuning $\delta$:  our effective bath responds slower when we are on resonance.  This is in sharp contrast to what is found for the cavity plus mirror system discussed above; there, the response time is always set by the cavity ring-down time, regardless of detuning \cite{Braginsky67, Metzger04, Harris05}.  

Finally, we wish to emphasize the smallness of $T_{eff}$ for the JQP process:  unlike the normal metal SET or tunnel junction, the effective temperature here is not set by the source-drain voltage, but rather by a much smaller scale, $\Gamma_a$.  As with the case of sequential quasiparticle tunneling, we see that the $T_{eff}$ of a mesoscopic conductor is by no means always set by the drain source voltage.  The effective temperature is rather a measure of the asymmetry between energy absorption and emission.  In the present case, the smallness of $\Gamma_a$ means that incoherent Cooper-pair tunneling can be used for substantial cooling of the resonator when $\delta > 0$; we give numerical estimates of the magnitude of this cooling effect in Sec. 4.3.

%Of more interest are the noise properties of this cycle. We again
%focus on the case of small $E_{J}/\Gamma$, and calculate the zero
%frequency charge noise associated with the fluctuations of the island
%charge $n$. We find simply:\begin{eqnarray}
% & S_{n}(0)=2\hbar^{2}\gamma_{CPT}(\delta)\cdot\frac{4\delta^{2}+\Gamma_{a}^{2}+4\Gamma_{a}\Gamma_{b}+8\Gamma_{b}^{2}}{\Gamma_{b}^{2}\left(4\delta^{2}+\Gamma_{a}^{2}\right)}\label{eq:JQPSQ}\end{eqnarray}
%On resonance, this expression reduces to $S_{n}(0)\propto\gamma_{CPT}/\Gamma^{2}$.
%Note that it is \emph{proportional} to current. We may understand
%this result as corresponding to telegraph noise in the limit where
%one of the rates is much smaller than the other. Here, we have telegraph
%noise between the charge states $n=0$ and $n=1$, and the incoherent
%Cooper-pair tunneling rate (from $n=0$ to $n=1$) is much slower
%than the quasiparticle transition rate from $n=1$ to $n=0$.

%%%%%%%%%%%%%%%%%%%%%%%%%%%%%%%%%%%%%%%%%%%%

\subsection{DJQP back-action}

We now turn to the back-action of a slightly more complex incoherent Cooper-pair tunneling
process, the Double Josephson Quasiparticle resonance (DJQP).  This cycle consists
of two incoherent Cooper-pair tunneling events in series, one in each junction (see Fig. \ref{fig:DJQP}b); its back-action properties (in the context of qubit measurements) were studied in
 \cite{Clerk02}.  The DJQP process is of interest as it 
occurs at a smaller drain-source voltage in the SET than the JQP process; as we will see, it also has
more pronounced back-action, and can have a greater cooling effect.

For definiteness, we consider the following cycle: a) Cooper-pair
tunnels from left, taking the island charge from $n=0$ to $n=2$;
b)Quasi-particle tunnels from island to right lead, taking $n=2$
to $n=1$; c)A Cooper-pair tunnels from the island to the right lead,
taking $n=1$ to $n=-1$ ; d) A quasiparticle tunnels onto the island
from the left lead, taking $n=-1$ to $n=0$ . At this stage, the
cycle repeats.  Note that there are two Cooper-pair tunneling events (one in each junction),
and two quasi-particle tunneling events (one in each junction). We
denote the first quasiparticle transition rate ($n=2$ to $n=1$) by $\Gamma_{a}$,
and the second by $\Gamma_{b}$. We also denote the first Cooper-pair
transition ($n=0$ to $n=2$) as ``A'', and the second transition
($n=1$ to $n=-1$) as ``B''. As with the JQP process, we have to worry
about the following voltage-dependent energies:
\numparts
\begin{eqnarray}
	\delta_{A} & = & 4E_{C}(1-\mathcal{N})-eV_{DS}\\
	\delta_{B} & = & -4E_{C}(0-\mathcal{N})-eV_{DS}\\
	E_{a} & = & 2E_{C}(3/2-\mathcal{N})+eV_{DS}/2\\
	E_{b} & = & -2E_{C}(-1/2-\mathcal{N})+eV_{DS}/2
\end{eqnarray}
\endnumparts
Here, $\delta_{A},$$\delta_{B}$ are the detunings of the two Cooper-pair
transitions, and $E_{a},E_{b}$ are the energies driving the two quasiparticle
transitions. The rates $\Gamma_{a},\Gamma_{b}$ of these quasiparticle
transitions are related to the corresponding driving energies by Eq.~
(\ref{eq:QPRate}) as before. Note that the center of the DJQP resonance
(where both $\delta_{A}$ and $\delta_{B}$ are 0) occurs when both 
$\mathcal{N} = 1/2$ and $eV_{ds} = 2E_{C}$.

An density-matrix approach identical to that used for the JQP may be used to describe
the transport and noise properties of the DJQP.  While this technique is valid
for arbitrary ratios $\Gamma_b / \Gamma_a$ and $E_J / \Gamma_a$, we will make some additional mild approximations to obtain results that may be
easily interpreted.  First, we again consider a SSET with identical junctions,
and note that at the center of the DJQP resonance, $\Gamma_a = \Gamma_b$.  As we are interested
in the behaviour near the resonance center, we will take $\Gamma_a = \Gamma_b = \Gamma$ throughout.
In addition, we will consider the limit $E_J \ll \Gamma$; this limit is approximately realized in many experiments, and
leads to a great simplification in the form of the resulting equations.  

In this limit, the Cooper-pair tunneling is the rate limiting step in the DJQP cycle; we find that the
average current is given by $3e$ times the series addition of the incoherent Cooper-pair tunneling rates for the ``A" and
``B" transitions:
\begin{eqnarray}
	I  	& = & 
			3 e \times 
				\left(
					\frac{1}{\Gamma_{CPT}(\delta_A)} +
					\frac{1}{\Gamma_{CPT}(\delta_B)}
				\right)^{-1}
		=
			\frac{3e}{2}\frac{E_{J}^{2}\Gamma}{(\hbar \Gamma)^{2}+2\delta_{A}^{2}+2\delta_{B}^{2}}
	\label{eq:DJQPIAvg}
\end{eqnarray}
The incoherent Cooper-pair tunneling rate $\Gamma_{CPT}(\delta)$ is defined in Eq.~ (\ref{eq:JQPAvgI}).

In calculating the back-action for the DJQP process, we make again make
the assumption that we can ignore the quasiparticle contribution to damping; as
discussed, near resonance, this contribution leads to terms which are higher order in the
dimensionless conductance $g$.  For the effective temperature, we again find a 
remarkably simple result:
\begin{eqnarray}	
	k_{B}T_{eff} 
		& = &
			\left[ 	\frac{1}{k_B T_{eff,A}} + \frac{1}{k_B T_{eff,B}} \right]^{-1} 
			\nonumber \\
		& = &
			\left[
				\frac{16\delta_A}{(\hbar \Gamma)^{2}+4\delta_A^{2}} +
				\frac{16\delta_B}{(\hbar \Gamma)^{2}+4\delta_B^{2}}
			\right]^{-1} 
			\nonumber \\
		& = &	
			\frac{  \left[ (\hbar \Gamma)^{2}	+	4\delta_{A}^{2}	\right]
				 \left[ (\hbar \Gamma)^{2}  +	4\delta_{B}^{2}	\right] }
			{
			16 \left[ \delta_{A}+\delta_{B} \right]
				\left[ (\hbar \Gamma)^{2}+4\delta_{A}\delta_{B} \right]}
		\label{eq:DJQPTeff}
\end{eqnarray}
We see that the effective temperature for the DJQP process (in the small $E_J$ limit) is simply given
by the series addition of the effective temperature for each individual Cooper-pair resonance;
the scale for the minimum effective temperature is again set by
$\Gamma$ (i.e. the lifetime of each of the two resonant states).  The equation for the 
back-action damping $\gamma$ is slightly more complicated:  
\begin{eqnarray}
	\frac{m \gamma}{A^2} & = &
		2 \frac{\delta_A + \delta_B}{E_J^2 \Gamma} 
			\left[
			\frac {
				( (\hbar \Gamma)^2 + 4 \delta_A^2)  ( (\hbar \Gamma)^2 + 4 \delta_B^2) 
					((\hbar \Gamma)^2 + 4 \delta_A \delta_B) 
			}{\left[ 
					(\hbar \Gamma)^2 + 2 \delta_A^2 + 2 \delta_B^2
			   \right]^3}
			   \right]
		\label{eq:DJQPgamma}
\end{eqnarray}
				  
\begin{figure}
\center{\includegraphics[width=16 cm]{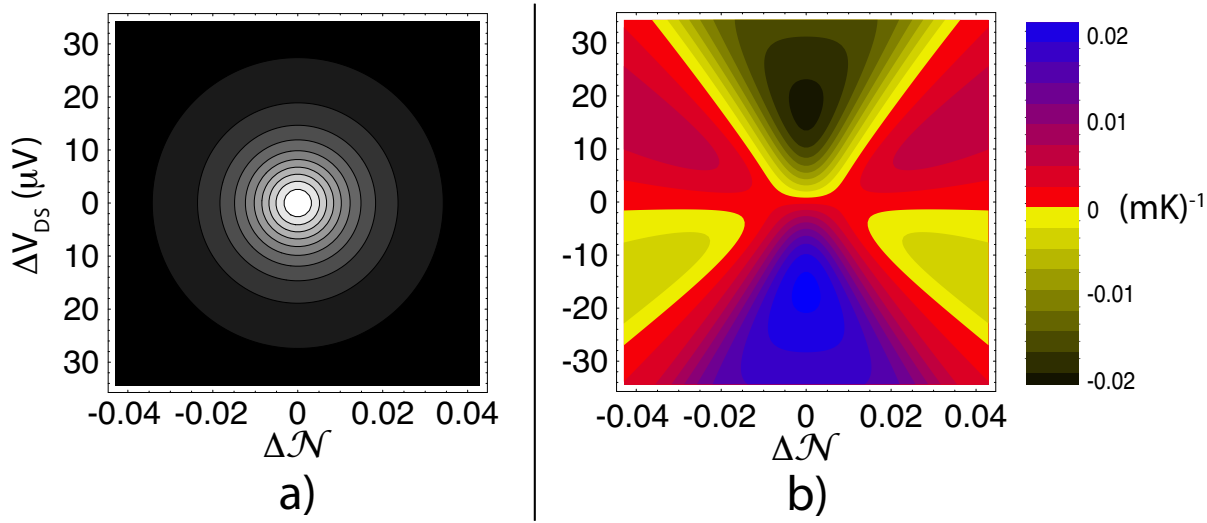}}
\vspace{-0.8cm}
\caption{\label{fig:DJQPTEff}
a) Contour plot (with equally spaced contours) of current near DJQP resonance, as a function of gate voltage detuning ($\Delta \mathcal{N} = \mathcal{N}-1/2$) and drain-source voltage detuning ($\Delta V_{DS} = V_{DS} - 2 E_C / e$) from the resonance center.  We have chosen typical SSET parameters $\Delta = E_C = 0.2 \textrm{meV}$ and $g=0.2$.  b) Contour plot of 
$\left(T_{osc}\right)^{-1}$,
the inverse of the oscillator's temperature (c.f. Eq.~ (\ref{eq:TOsc})), 
for the same range of $\Delta V_{DS}$ and $\Delta \mathcal{N}$.  We have taken typical oscillator parameters $\Omega/(2 \pi) = 27 \textrm{MHz}$,
$Q_0 = 10^4$ (i.e. intrinsic oscillator quality factor),
 $d = 300 \textrm{nm}$ (i.e. oscillator-SET island distance),
 $C_{osc} = 0.1 C_{\Sigma}$ 
 and $V_{osc} = 10 \textrm{V}$ (c.f. Eq.~ (\ref{eq:SETA})); we have also chosen a bath temperature of $T_0 = 500 \textrm{mK}$.  Yellow indicates unstable regions where the oscillator temperature is negative.  In the regions of cooling (blue), the lowest achievable oscillator temperature is less than $50 \textrm{mK}$, a factor of ten lower than the oscillator's temperature at zero coupling to the SET.
}
\end{figure}

Similar to the situation with JQP, the effective temperature (and damping) 
can become negative if the {\it net} 
preference of the two resonances is to emit energy to the oscillator
(i.e. the series addition of $T_{eff,A}$ and $T_{eff,B}$ must be negative).
To see exactly where negative damping and temperature occur in terms of SSET operating points (i.e. $V_{ds}$ and $\mathcal{N}$),
it is useful to re-write the expression for $T_{eff}$.  We define: 
\numparts
\begin{eqnarray}
4E_{C}\mathcal{N} & = & 4E_{C}(1/2+\delta\mathcal{N})=2E_{C}+\Delta_{\mathcal{N}}\\
eV_{DS} & = & 2E_{C}+e\delta V_{DS}=2E_{C}+\Delta_{V}
\end{eqnarray}
\endnumparts
$\Delta_{\mathcal{N}}$  and $\Delta_{V}$ 
represent the energy detuning that result when
we move, respectively, $\mathcal{N}$ or $V_{DS}$ from the center of the
resonance. With this notation, we have:\begin{eqnarray}
k_{B}T_{eff} & = & \frac{1}{-32\Delta_{V}}\cdot\frac{\left[\Gamma^{2}+4(\Delta_{V}+\Delta_{\mathcal{N}})^{2}\right]\left[\Gamma^{2}+4(\Delta_{V}-\Delta_{\mathcal{N}})^{2}\right]}{\Gamma^{2}+4\left[(\Delta_{V})^{2}-(\Delta_{\mathcal{N}})^{2}\right]}\end{eqnarray}
For small detunings from the center of the resonance ($\Delta_{V},\Delta_{\mathcal{N}}\ll\Gamma$),
we see that the sign off the effective temperature is completely determined
by the value of the drain-source voltage. If $V_{DS}$ is set to be
below the center of the DJQP resonance, $\Delta_{V}$ is negative,
and hence both $T_{eff}$ and the damping $\gamma$ are positive.
In this case, the net preference of the two resonances is to absorb energy from the oscillator.
In contrast, if the
drain-source voltage is tuned to be higher than resonance, we get
both negative temperature and negative damping. Now, the DJQP process
can be brought closer to resonance by \emph{emitting}
energy to the oscillator.  Finally, note if the drain source voltage
is exactly at the center of the resonance (i.e. $\Delta_V=0$), $T_{eff}$ tends to infinity,
meaning that there is no damping, and no asymmetry between absorption
and emission; this holds regardless of how much one moves the gate
voltage $\mathcal{N}$ off-resonance. In this special case, one Cooper-pair
transition always prefers to absorb energy, the other to emit; the
net result is no preference between absorption and emission.  
In Fig. \ref{fig:DJQPTEff} we show a contour plot of $T_{eff}$ for DJQP using typical
device parameters, and compare it against a plot of the current. 

It is instructive to compare the magnitude of detector-induced
damping for the JQP and DJQP processes in the small $E_J$ limit we consider.
In both cases, the minimum value of $|T_{eff}|$ is $\propto \Gamma$, where
$\Gamma$ is the quasiparticle tunneling rate.  However, the corresponding 
damping will be much larger in magnitude for DJQP versus JQP:  
comparing Eqs. (\ref{eq:JQPgamma}) and 
(\ref{eq:DJQPgamma}), we see that the back-action damping in the DJQP case is enhanced over 
that at JQP by a large factor $(\Gamma / E_J)^4$.  
The reason for this is simple to understand, once we recall that $\gamma \propto S_n(0) / T_{eff}$, where
$S_n(0)$ is the zero-frequency charge noise of the SSET island.  For both JQP and DJQP, this noise has a telegraph-noise form in the small $E_J$ limit.  For JQP, we have effective telegraph noise between
the $n=0$ and $n=2$ charge states, where one rate 
(the quasiparticle transition rate $\Gamma$) is much larger than the other (the incoherent Cooper-pair tunneling rate $\Gamma_{CPT}$).  As a result, $S_n(0) \propto \Gamma_{CPT} / \Gamma^2$.  In contrast, for DJQP, we
have effective telegraph noise between the $n=0$ and $n=1$ charge states; the two telegraph rates are both incoherent Cooper-pair tunneling rates (i.e. the ``A" and ``B" transitions), and are roughly equal.  In this case $S_n(0) \propto 1/ \Gamma_{CPT}$.  Thus, as both processes have similar effective temperatures, 
the enhanced charge noise of DJQP explains its enhanced damping.  This
enhancement of back-action damping should make it easier to see back-action effects of DJQP in an experiment.  

Finally, consider the response time $\tau_{resp}$ of the effective bath presented by DJQP:  
it is given by:
\begin{eqnarray}
	\frac{1}{\tau_{resp}} &=& 
		\Gamma_{CPT}(\delta_A) + \Gamma_{CPT}(\delta_B)
		\nonumber \\
		& = &	
		\frac{2 E_J^2}{\hbar^2 \Gamma} \times
			\frac				{
				\left( \hbar \Gamma \right)^2 
				\left[
					(\hbar \Gamma)^2 + 2 \delta_A^2 + 2 \delta_B^2
				\right]
				}
				{
				\left[ (\hbar \Gamma)^2 + 4 \delta_A^2 \right]
				\left[ (\hbar \Gamma)^2 + 4 \delta_B^2 \right] }
		\label{eq:DJQPtau}
\end{eqnarray}
The average response rate, $1/\tau_{rep}$, is simply given by the sum of the
rates of the two 
Cooper-pair tunneling events in the cycle.  In the small $E_J$ limit we are considering, this timescale is much longer than what was found for JQP (c.f. Eq.~(\ref{eq:JQPtau})),
where one always had $\tau_{resp} \sim \frac{1}{\Gamma}$.  

As a final note, we should remark that none of the approximations
used here (small $E_{J}$, no quasiparticle contribution to damping)
are necessary to the method used.  We have 
made these approximations only so that we could give a clear
heuristic picture of the physics underlying the back-action effects
found.

\subsection{Typical Cooling Temperatures}

Eq.~ (\ref{eq:TOsc}) tells us that if the back-action damping dominates the intrinsic, non-back-action damping of the oscillator (i.e. $\gamma \gg \gamma_0$), then it should be possible to cool the oscillator to the effective temperature of the SET, even if this is much lower than $T_0$ (i.e. the temperature of the oscillator when it is decoupled from the SET).  To achieve $\gamma \gg \gamma_0$, one should start with an oscillator with a high intrinsic quality factor, and engineer structures where the SET-oscillator coupling is relatively strong (e.g.. have the SET island be in close proximity to the oscillator)
Note that this condition does not imply any violation of the weak-coupling assumptions of our treatment:  it is entirely possible to have the coupling $A$ be weak-enough that linear response is still valid, yet strong enough that $\gamma \gg \gamma_0$.

To give an idea of the kind of cooling that should be achievable using incoherent Cooper-pair tunneling, it useful to give some numerical estimates using typical SET parameters.  We assume $E_C = \Delta = 0.2 meV$, in agreement with the devices used in Ref. 
\cite{Lehnert03}.  For the JQP, we saw that the minimum possible positive value of $T_{eff}$ was given by $\Gamma_a/4$.  Choosing $V_{DS} = 2 \Delta + E_C$ (the threshold voltage for the JQP process), we find for these parameters:
\begin{equation}
	T_{eff} \Big |_{min} \simeq g \times 350 mK
\end{equation}
Thus, for $g=0.2$ (i.e. a junction resistance of $130 k \Omega$ ), the minimum effective temperature is approximately $70 mK$; again, this is the minimum temperature we could cool our oscillator to using the JQP process.  Note that this $T_{eff}$ is much lower than the scale set by the voltage: $eV_{DS} / k_B \simeq 7.0 K$.

Turning to the DJQP, and now assuming the typical condition $E_J < \Gamma$, we see from Eq.~(\ref{eq:DJQPTeff}) that the minimum possible positive value of $T_{eff}$ is $\Gamma/8$, a full factor of two smaller than for the JQP process.  If we again take $E_C = \Delta = 0.2 meV$, we find:
\begin{equation}
	T_{eff} \Big |_{min} \simeq g \times 125 mK
\end{equation}
For systems with $E_J < \Gamma$, cooling should be much easier to achieve using DJQP.  Not only is the $T_{eff}$ lower, but, as already discussed, the back-action damping due to DJQP is larger than that of JQP by the large factor  $(\Gamma / E_J)^4$.

As a final caveat, the above estimates assume that the only source of broadening of the Cooper-pair tunneling resonances is quasiparticle tunneling.  In reality, additional effects could further broaden these resonances and increase the minimum achievable temperature.  Perhaps the most important of these will be voltage fluctuations associated with the environmental impedance seen by the SET island.  These effects will ultimately limit how much one can lower $T_{eff}$ by simply lowering $g$.

\subsection{Quantum limited measurement with incoherent CPT}

\begin{figure}
	\center{
		\includegraphics[width=14.0 cm]{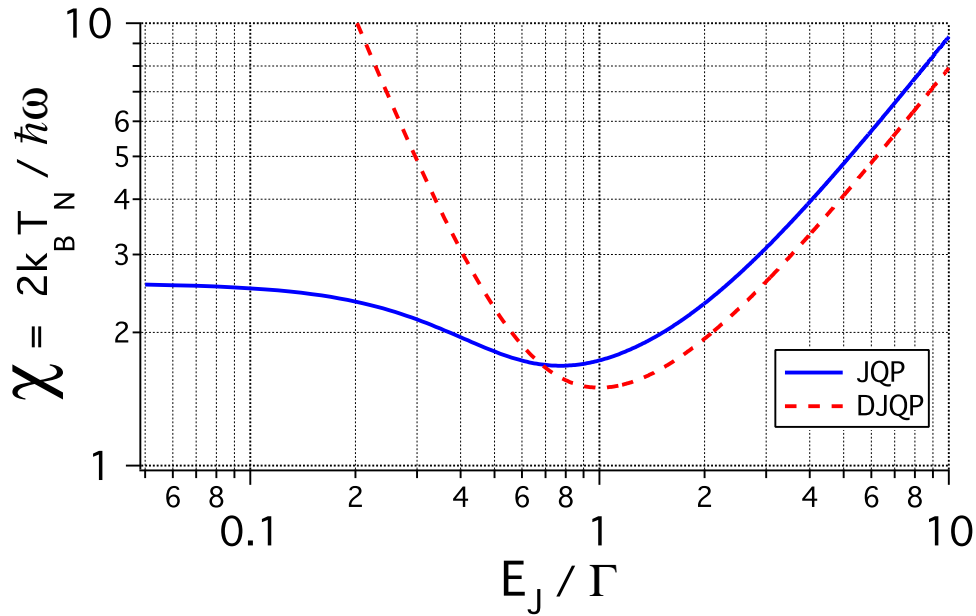} \caption{Noise temperature for the JQP and DJQP processes (scaled by $\hbar \Omega / 2$) versus $E_J / \Gamma$, where $E_J$ is the Josephson energy, and $\Gamma$ is the quasiparticle tunneling rate.  We have assumed an optimal choice of coupling strength $A$, identical SET junctions, and that both quasiparticle tunneling rates in each cycle are the same.
		\label{fig:QLPlot}}}
\end{figure}

There is an additional important point that needs to be made about incoherent 
Cooper-pair tunneling processes:  unlike sequential tunneling in a normal SET, these processes 
may be used to do near quantum-limited displacement detection \cite{Clerk02}.  As discussed extensively in \cite{Clerk04b}, reaching the quantum limit requires a detector which has a minimal amount of back-action noise relative to the amount of information it provides.  The measure of this relative back-action is the noise temperature $T_N$ of the detector, which gives a measure of how much noise is added to the signal by the detector.  The quantum limit is that $k_B T_N$ cannot be any smaller than $\hbar \Omega / 2$, where $\Omega$ is the oscillator's frequency \cite{Caves82}.
We are interested in the usual limit where the detector noise correlators are frequency independent on frequency scales relevant to the oscillator (i.e. $\Omega, \gamma$).
After optimizing the strength of the coupling $A$ (to balance back-action and intrinsic shot noise contributions to the total noise), one has \cite{Clerk04b}:
\begin{equation}
	\chi \equiv
	\frac{k_B T_N}{\hbar \Omega / 2} = \sqrt{
		\frac{4 S_I(0) S_F(0)}{ \left(  \hbar dI/dx  \right)^2 }
		}   \geq 1
\end{equation}
Here, $S_F = A^2 S_n$ is the back-action force noise spectrum of the detector, $S_I$ is the zero frequency current noise of the SSET, and $d I / dx = -(dI/d \mathcal{N}) \frac{2 A}{E_C}$ is the gain of 
our detector:  how strongly does the average current respond to changes in the oscillator's position.

Thus, achieving the quantum-limit of displacement detection requires that $\chi$ approach one
	\footnote{
		Note that if our detector was used to do a weak QND measurement of a
		qubit, $1/\chi^2$ represents the ratio between the rate at which information
		is acquired by the measured versus the back-action dephasing rate.}.
For a normal-metal SET in the sequential tunneling regime, $\chi$ is proportional to 
$1/g^2$, where $g$ is the dimensionless conductance of the junctions.  By assumption, $g$ is a small parameter-- this is what allows us to use perturbation theory in the tunneling!  One thus has that sequential tunneling in a normal SET is far from being at the quantum limit; there is considerable excess back-action noise, far beyond what is required by quantum mechanics.  In contrast, for JQP and DJQP, $\chi$ is order 1, and is independent of $g$ \cite{Clerk02}. 
Using the same density matrix approach used to calculate the charge noise, one can calculate the zero frequency current noise of both JQP and DJQP \cite{Choi01, Clerk02, Clerk03b}.   Shown in Fig. \ref{fig:QLPlot} is the reduced noise temperature $\chi$  for JQP and DJQP, evaluated as a function of the ratio $E_J / \Gamma$, and for a bias point which yields a maximum gain.  Note that the ratio $E_J / \Gamma$ can in principle be controlled if one can control the ratio of $E_C$ to $\Delta$ when fabricating the SET.  We see when that $E_J \simeq \Gamma$, both DJQP and JQP come close to reaching the quantum limit, with the DJQP process doing slightly better.  In the more usual case where $E_J \ll \Gamma$, the JQP process is much better at approaching the quantum limit than the DJQP process; this is because of the relative suppression of charge noise associated with JQP when $E_J \ll \Gamma$.

\section{Negative Damping and Strong Electro-mechanical Feedback}

In this last section, we describe in more detail the interesting 
negative-damping instability that can be brought about by incoherent Cooper-pair tunneling in a SET plus mechanical resonator system.  As discussed, the negative back-action damping associated with either the JQP or DJQP transport cycles can, for a sufficiently strong coupling, drive both the
{\it total} oscillator damping and its temperature 
$T_{osc}$ (as given by Eq.~ (\ref{eq:TOsc})) below zero (e.g. see the yellow regions in Fig. \ref{fig:DJQPTEff}b).  
In this regime, the SET will continually dump energy into the oscillator, causing the amplitude of the oscillator's motion to continually increase.  Eventually, this amplitude will become so large that there will be an effective strong coupling between the oscillator and the SET-- the motion of the oscillator will strongly effect the dynamics of the SET.  In this regime, the weak coupling approximations we have been making clearly break down.  We nonetheless wish to describe the properties of the eventual stationary state achieved in this regime.  Note such negative-damping instabilities are also well known in the analogous optical cavity system \cite{Braginsky70, Braginsky77, Marquardt05}.

To deal with the feedback occurring in this effective strong coupling regime, 
we will make the 
usual simplifying assumption that the oscillator is much slower than the SET: $\Omega \ll \Gamma, \Gamma_{CPT}$; this condition is usually more than satisfied in experiment.  We can then make use of this separation of timescales in manner somewhat analogous to the usual Born-Oppenheimer approximation.  As in our original weak-coupling description, at each instant in time the SET acts as an effective thermal bath on the oscillator, and is characterized by a fluctuating back-action force with spectral density $S_{\delta f}$ and a damping rate $\gamma$; the oscillator is thus still described by the Langevin equation Eq.~ (\ref{eq:Langevin}).  Now, however, the parameters of the bath will depend on the instantaneous position of the oscillator:  this is how we deal with the strong-coupling aspect of the problem. Thus, the spectral density of the back-action force $\delta f(t)$ appearing in Eq.~ (\ref{eq:Langevin}) will be x-dependent (i.e. 
$S_{\delta f} = S_{\delta f}[x]$), as will the damping coefficient $\gamma$ (i.e. $\gamma = \gamma[x]$).  The $x$ dependence of both these quantities arises completely through their dependence on the SET dimensionless gate voltage $\N$; we can absorb the coupling Hamlitonian of Eq.~ (\ref{eq:Hint}) into the SET charging energy Hamiltonian (c.f. Eq.~ (\ref{eq:HC})) by writing:
\begin{eqnarray}
	\N[x] = \N[0] + \frac{A}{2 E_C} x = \frac{C_g V_g}{e} + \frac{A}{2 E_C} x.
	\label{eq:NNx}
\end{eqnarray}
Again, the key assumption here is that the SET response to changes in $x$ is much faster than the evolution of $x$ itself.

As we show in what follows, the net result of this approach is that we may describe our coupled system with a classical Fokker-Planck equation where {\it both} the diffusion and damping constants have an explicit $x$ dependence.  Heuristically, for large oscillator amplitude, the Cooper-pair detuning $\delta$ and hence the back-action damping will depend on the position of the oscillator; in the unstable regime of interest, the stationary state of the oscillator will correspond to an amplitude of oscillation so large that the oscillator experiences equal amounts of positive and negative damping during one period of its motion.  More quantitatively, we find that the stationary state of oscillator is naturally characterized by an {\it energy-dependent temperature}; for a variety of regimes, the energy distribution of the oscillator is a gaussian, and hence highly non-thermal.  Note that a similar approach to potential strong-feedback behaviour was used in Ref. \cite{Blanter05}; however, the particular NEMS systems studied in that work did not exhibit any negative damping, and thus there was no effective strong-coupling regime.  

Our starting point is thus the Langevin equation of Eq.~ (\ref{eq:Langevin}) with $x$ dependent damping and back-action force terms.
%For simplicity, we will focus on the JQP process; similar results are obtained for DJQP.  As usual, we will consider a SSET with identical junctions, and take the dimensionless conductance $g$ to be sufficiently small that we can neglect the quasiparticle contribution to damping.  In this case, all feedback effects of the oscillator's motion on the SET occur via the $x$-dependence of the Cooper-pair detuning:
%\begin{eqnarray}
%	\delta[x] = \delta[x=0] - 2 A x
%\end{eqnarray}
As we are interested in the limit where the oscillator is much slower than the source of the back-action noise, we can treat the latter as being white.  We may then convert our Langevin equation to a Fokker-Planck equation for the oscillator's phase-space density $w(x,p;t)$ in the usual way \cite{Risken96}; we obtain:
\begin{eqnarray}
	\frac{\partial}{\partial t} w & = &
		\left[
			-\frac{p}{m} \frac{\partial}{\partial x} +
			\frac{\partial}{\partial p} \left( m \Omega^2 x + 
				\left(\gamma_0 + \gamma[x]) p \right) \right)
		\right] w
		+ \left(D_0 + D[x]\right) \frac{\partial^2}{\partial p^2}w.
		\nonumber \\
		&&
	\label{eq:FokkerPlank}
\end{eqnarray}
Here $\gamma_0$ is the intrinsic (SET-independent) damping of the oscillator, $D_0  = m \gamma_0 k_B T_0$ describes momentum diffusion due to the equilibrium bath, and the $x$-dependent diffusion constant $D[x]$ is determined from 
back-action force spectrum: 
\begin{eqnarray}
	D[x] = \frac{1}{2} S_{\delta f}[\omega=0;x] = m \cdot \gamma[x] \cdot k_B T_{eff}[x] 
\end{eqnarray}
In the case of JQP, $\gamma(x)$ and $T_{eff}[x]$ are given respectively by Eqs. (\ref{eq:JQPgamma}) and (\ref{eq:JQPTeff}) with the substitution $\N \rightarrow \N[x]$;
for DJQP, one would use Eqs. (\ref{eq:DJQPgamma}) and (\ref{eq:DJQPTeff}).
Note that there is no ambiguity in interpreting Eq.~(\ref{eq:Langevin}) with an $x$-dependent back-action noise spectrum: both the Ito and Stratonovitch interpretations of this stochastic differential equation yield the same Fokker-Planck equation.  Also note that we are neglecting the $x$-dependence of the conservative part of the average back-action force; this is justified in the experimentally relevant limit of a high intrinsic quality factor $Q_0$.  In this limit, one can have the magnitude of back-action damping be comparable or greater than the intrinsic damping of the oscillator, while at the same time have the back-action spring-constant modification $\Delta k$ be much much smaller than $m \Omega^2$.

To make further progress, we consider the experimentally-relevant weak-damping limit of Eq.~ (\ref{eq:FokkerPlank}) (i.e. $\gamma_0, |\gamma[x]| \ll \Omega$), and follow the approach of Kramers \cite{Kramers41} (and more recently, Blanter et. al \cite{Blanter05}).  We first re-write $x$ and $p$ in terms of the oscillator's energy $E$ and phase $\theta$ 
($x = \sqrt{2 E / k} \sin \theta$, $p = \sqrt{2 m E} \cos \theta$), 
and then convert Eq.~ (\ref{eq:FokkerPlank}) into an equation for $w(E,\theta)$.
Working to lowest order in the dissipative terms (both damping and diffusion), this equation can then be recast as an equation for the oscillator energy distribution $w(E) = \int_0^{2 \pi} d \theta \phantom{\cdot} w(E,\theta)$; one uses the fact that in the absence of any dissipation $w_0(E,\theta) = w_0(E) / 2 \pi$.  After some algebra, this procedure yields:
\begin{eqnarray}
	\frac{d}{dt} w(E;t) = \frac{\partial}{\partial E} E \left(
		\gamma_0 + \gamma(E) + \frac{D_0 + D(E)}{m} \frac{\partial}{\partial E}
		 \right) w(E;t)
	\label{eq:pE}
\end{eqnarray}
The energy-dependent back-action damping and diffusion constants here are defined as:
\begin{eqnarray}
	\gamma[E] & = & 2 \int_0^{2 \pi} \frac{d \theta}{2 \pi} 
		\gamma \left[
			x = \sqrt{2E/k} \sin \theta
			\right]  \cdot \cos^2 \theta 
			\label{eq:GammaE}
		\\
	D[E] & = & 2 \int_0^{2 \pi} \frac{d \theta}{2 \pi}
		D \left[
			x = \sqrt{2E/k} \sin \theta
		\right] \cdot \cos^2 \theta 
\end{eqnarray}		
As could have been expected, they are given by averaging $\gamma[x]$ and $D[x]$ over the phase of the oscillator at fixed amplitude; what is perhaps more surprising is that one must take a weighted average, with a weighting factor which is proportional to $p^2$.
Note that at $E=0$, $\gamma(E)$ and $D[E]$ coincide respectively with the back-action damping and diffusion constant in the weak-coupling (i.e. zero feedback) theory.
	
It is now trivial to solve for the stationary energy distribution of the oscillator.  From Eq.~ (\ref{eq:pE}), we have:
\begin{eqnarray}
	w(E)_{stat} = N \exp \left(
		- \int_0^{E} \frac{d E'}{k_B \widetilde{T}_{osc}[E']} 
		\right)
	\label{eq:pEstat}
\end{eqnarray}
where $N$ is a normalization constant, and the energy-dependent effective temperature 
$\widetilde{T}_{osc}(E)$ is given quite naturally by:
\begin{eqnarray}
	 k_B \widetilde{T}_{osc}(E) \equiv \frac{D_0+ D[E] }
	 	{m\left( \gamma_0 + \gamma[E]\right)}
\end{eqnarray}
We see that the stationary state is characterized by a generalized Boltzmann distribution having an energy-dependent effective temperature.  In the weak-coupling, no negative damping case, we can neglect the energy dependence of $k_B \widetilde{T}[E]$, and
Eq.~ (\ref{eq:pEstat}) coincides with a thermal distribution and with the results of the weak-coupling theory (c.f. Eq; (\ref{eq:TOsc})).  In the strong-feedback regime of interest, the total damping $(\gamma_0+\gamma(E))$ is negative at $E=0$; as we increase $E$, the motion of the oscillator smears out the back-action contribution to the damping, and for very large $E$, the total damping tends to $\gamma_0>0$.  There will thus be a critical $E=E_0$ where  $(\gamma_0 + \gamma(E_0))$ will pass through zero.  It follows directly from Eq.~ (\ref{eq:pEstat}) that $w(E)$ will have a maximum at $E_0$; moreover, in the vicinity of $E_0$, $w(E)$ will look gaussian:
\begin{eqnarray}
	w(E)_{stat} & \simeq& 
		N \exp \left (
		 - 
		\frac{\left( E-E_0 \right)^2 } {2 \sigma^2} 
			 \right) \\
	\sigma^2 & = & 
		\frac{D_0 + D[E_0]}{2 m} 
		\left[\frac{d \gamma}{d E} \Bigg|_{E=E_0} \right]^{-1}
	\label{eq:pEsigma}
\end{eqnarray}
We thus find that in the strong-feedback regime, the oscillator's energy distribution function has a maximum at a non-zero energy, in sharp contrast to an equilibrium distribution.  The energy at which the maximum occurs corresponds, as expected, to an oscillator amplitude large enough that the oscillator experiences zero average damping during each period of its motion.  The width of the distribution near this maximum is set by {\it both} the back-action noise and the energy-sensitivity of the damping via Eq.~ (\ref{eq:pEsigma}).

\begin{figure}
	\center{
	\label{fig:JQPGamma}
	\includegraphics[width=12.0 cm]{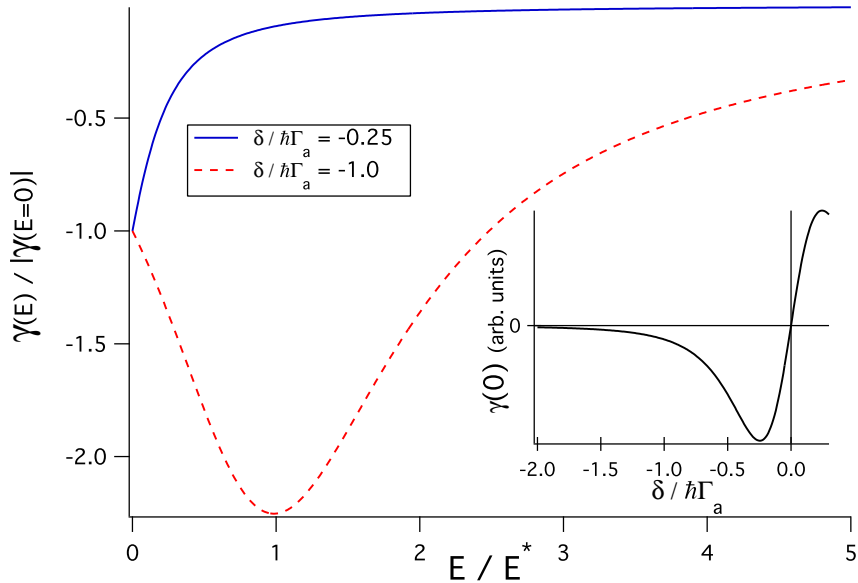} 
	\caption{Energy-dependent back-action damping of the JQP process, 
	c.f. Eq. (\ref{eq:GammaE}), plotted as a function of energy, and
	using SSET parameters identical to those used in Fig. 2.
	 The characteristic
	scale for feedback effects, $E^*$, is given in Eq. (\ref{eq:EStar}).
	Note that the chosen scaling makes this plot independent of the coupling
	strength $A$.  The red curve and blue curves correspond to two 
	different Cooper-pair detunings
	$\delta$ (i.e. SSET operating points) as labeled; note that $\gamma(E)$ can
	have a non-monotonic dependence on energy.  The inset shows the dependence
	of the back-action damping at zero oscillator energy 
	(c.f. Eq. (\ref{eq:JQPgamma})) on the Cooper-pair detuning.}}
\end{figure}

We now apply the above results, which are quite general, to the specific 
instability brought on by incoherent Cooper-pair tunneling.  We first consider the case of  an oscillator-SSET system operated in the negative damping regime near the JQP resonance; as usual we consider the limit of small dimensionless junction conductances, and neglect the quasiparticle contribution to the back-action damping.  In this limit, all the feedback effects of the oscillator on the SET will be due to the dependence of the Cooper-pair detuning $\delta$  (c.f. Eq. (\ref{eq:DeltaDefn})) 
on the oscillator position $x$.  The characteristic
energy scale $E^*$ for feedback effects will correspond an oscillator amplitude large enough that the corresponding oscillations in $\delta$ are equal to the resonance width, $\hbar \Gamma_a$.  From Eqs. (\ref{eq:NNx}) and (\ref{eq:DeltaDefn}), one has:
\begin{eqnarray}
	E^* = \left(
			\frac{\hbar \Gamma_{a}}
			{2 A \sqrt{\frac{2}{k}}  }
		\right)^2
		\label{eq:EStar}
\end{eqnarray}
$E^*$ sets the characteristic scale for variations in $\gamma[E]$ and $D[E]$.
Shown in the inset of Fig. 4  is the back-action damping $\gamma$ at zero energy (i.e. ignoring feedback effects),  c.f. Eq.~ (\ref{eq:JQPgamma}), as a function of $\delta$.  The main plot shows the energy dependence (scaled by $E^*$) of the the damping for two choices of $\delta_0 = \delta[x=0]$.  Note that $\gamma(E)$ can have a non-monotonic dependence on $E$.  

\begin{figure}
	\center{
	\label{fig:JQPw}
	\includegraphics[width=16.0 cm]{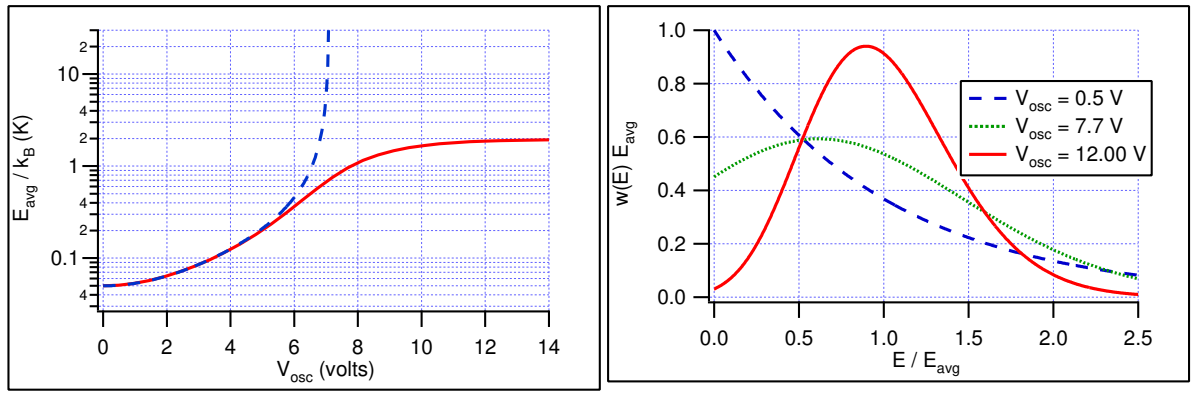} 
	\caption{Effects of JQP back-action in the negative-damping regime.  
	We have taken a bath temperature
	$T_{0}=50$ mK, an intrinsic oscillator quality factor $Q_0 = 10^5$, 
	and a SSET operating point near the JQP resonance
	which maximizes the effect of the back-action damping ($\delta = -0.25 \hbar \Gamma_a$).  All other parameters
	are the same as in Fig. 2.  Left panel:  Average energy of the oscillator
	$E_{avg}$  as a function of the coupling voltage $V_{osc}$; the dashed curve is the predictions of the weak-coupling theory, c.f. Eq.~ (\ref{eq:TOsc}), while
	the red curve is from the strong-feedback theory.  The total oscillator
	damping becomes negative at $V_{osc} = 7.1$ V.  
	Right panel: Oscillator energy distribution $w(E)$ as determined by Eq.~ (\ref	{eq:pE}).  As $V_{osc}$ becomes large enough for the strong-feedback
	effect to take hold, the energy distribution becomes gaussian-like. }}
\end{figure}

\begin{figure}
	\center{
	\label{fig:JQPw2}
	\includegraphics[width=16.0 cm]{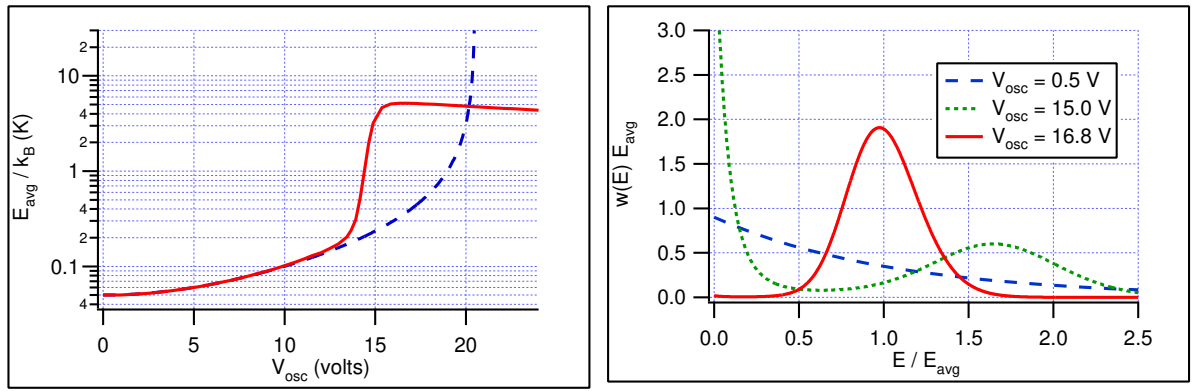} 
	\caption{Effects of JQP back-action in the negative-damping regime.  
	We use the same parameters as Fig.\ref{fig:JQPw}, but now choose
	an SSET operating point near the JQP resonance
	which yields a non-monotonic $\gamma(E)$. ($\delta = -\hbar \Gamma_a$). 
	Left panel:  Average energy of the oscillator
	$E_{avg}$  as a function of the coupling voltage $V_{osc}$, labeled as in 
	Fig.\ref{fig:JQPw}.  Note that the strong-feedback effect sets 
	at $V_{osc}$ much lower than what is needed to make the weak coupling
	theory diverge.      
	Right panel: Oscillator energy distribution $w(E)$ as determined by Eq.~ (\ref	{eq:pE}).}}
\end{figure}

While measuring the full distribution $w(E)$ of the oscillator's energy may be quite difficult, the average oscillator energy $E_{avg}$ can be obtained directly from experiment.  In the experiment of LaHaye {\it et. al} \cite{LaHaye04},
this quantity was obtained by extracting $\langle x^2 \rangle$ from the current noise of the SET, and then invoking the equipartition theorem; a similar approach could be used
in the strong-feedback regime discussed here
	\footnote{As the stationary state is characterized by a flat distribution of 
		the oscillator phase $\theta$,
		one again has the relation $E_{avg} /k  = \langle x^2 \rangle$, as would
		hold in equilibrium}.
In Fig. \ref{fig:JQPw} a), we plot $E_{avg}$ as a function of the coupling voltage $V_{osc}$; we use typical device parameters (see figure caption), and  choose $\delta = -0.25 \hbar \Gamma_a$ to maximize $| \gamma(E=0) |$ (i.e. same as the blue curve in Fig. 4.).  For small $V_{osc}$, the intrinsic damping of the oscillator dominates the back-action damping, and there is no strong feedback effect.  The oscillator is in a thermal state, with a temperature $T_{osc}$ given by Eq.~ (\ref{eq:TOsc}), and $E_{avg} = k_B T_{osc}$.  As we increase the coupling voltage, the back-action damping increases in magnitude (i.e. becomes more negative),
and $E_{avg} = T_{osc}$ correspondingly increases.  
At a critical voltage $V_{osc} = 7.1$ volts, the total damping becomes negative and the weak-coupling expression for $T_{osc}$ diverges.  The strong-feedback effect takes hold here, and we see that average energy
of the oscillator remains finite.  There is 
also a corresponding change in the shape of the energy distributions, as can be seen in Fig. \ref{fig:JQPw} b).  Note that for a large range of $V_{osc}$, $E_{avg}$ remains approximately constant.  This is the result of two competing tendencies:  increasing $V_{osc}$ increases the overall magnitude of the back-action damping, but also increases the sensitivity of the Cooper-pair detuning $\delta$ to $E$.

In Fig. \ref{fig:JQPw2}, we plot the same quantities as \ref{fig:JQPw}, but now for $\delta = -\hbar \Gamma_a $.  In this case, the non-monotonic nature of $\gamma(E)$ leads to energy distributions $w(E)$ which have two local maxima.  Also note that in this case, the 
strong feedback effect sets in well before the weak-coupling expression for $T_{osc}$ diverges.

\begin{figure}
	\center{
	\label{fig:DJQPw}
	\includegraphics[width=16.0 cm]{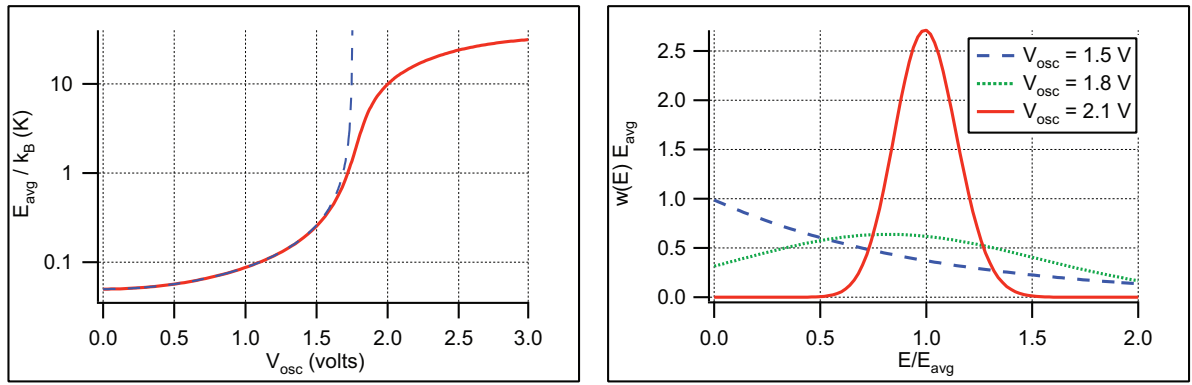} 
	\caption{Effects of DJQP back-action in the negative-damping regime.  
	We have taken a bath temperature
	$T_{0}=50 mK$, an intrinsic oscillator quality factor $Q_0 = 10^4$, 
	and a SSET operating point near the DJQP resonance (
	$\Delta \N = 0, \Delta V = \hbar \Gamma / 2)$.  All other parameters
	are the same as in Fig. 2.  Left panel:  Average energy of the oscillator
	$E_{avg}$  as a function of the coupling voltage $V_{osc}$; the dashed curve is the predictions of the weak-coupling theory, c.f. Eq.~ (\ref{eq:TOsc}).  The total damping becomes negative when the coupling voltage $V_{osc} = 1.75$ volts.  Right panel: Oscillator energy distribution $w(E)$ as determined by Eq.~ (\ref{eq:pE}).  At $V_{osc} = 
	1.5$ volts, we still have a thermal distribution; increasing the coupling voltage takes
	one into the unstable regime and results in an almost gaussian distribution of energy.  }}
\end{figure}

Finally, in Fig \ref{fig:DJQPw} we plot results for an oscillator SSET system operated in the negative-back action damping regime near the DJQP resonance.  We have again chosen realistic parameter values (see figure caption).  We see that the feedback effect is stronger here; as discussed, this is a consequence of the enhanced back-action damping of DJQP versus JQP.  Note that for both the JQP and DJQP strong feedback results presented here, the maximum amplitude of the oscillator is still much smaller than the oscillator-SET gap $d$, and the corresponding oscillation of the Cooper-pair detuning is at most order $\Gamma$.  Thus, these results do not violate our assumption of a linear coupling Hamiltonian, and do not involve SSET physics far from the JQP or DJQP resonance.

There are a number of other interesting issues related to the strong feedback regime discussed here.  In particular, what are the dynamics of the oscillator's energy in this regime?  As the total damping vanishes at the most probable energy $E_0$, one expects very slow dynamics.  How does this then manifest itself in the output noise of the SET?  It would also be interesting to investigate systems where the SET is not infinitely fast compared to the oscillator; in this case, one could expect similar kinds of multistability effects found in optical cavity systems \cite{Marquardt05}.  We hope to address these issues in the near future.

\section{Conclusions}

In this paper, we have discussed various aspects of back-action physics in SET-oscillator NEMS using a quantum-noise approach.  In the sequential tunneling regime, we derived general expressions for the effective temperature, damping rate, and bath response time that are valid regardless of the particular system (electrons in a normal SET, quasiparticles in a superconducting SET, electrons in a quantum dot).  In the regime of incoherent Cooper-pair tunneling, we described the back-action of both the JQP and DJQP processes, and demonstrated that both could be used for substantial oscillator cooling, as well as near quantum-limited measurement.  We also discussed the strong analogy between these processes and an optical cavity with a moveable mirror.  Finally, we discussed the regime of strong electro-mechanical feedback that can arise when the back-action damping becomes negative, and showed that the stationary state in this regime is naturally characterized by an energy-dependent temperature.

This work was supported by NSERC under grant RGPIN-311856, by the Canada Research Chairs programs, and by the FQRNT under grant 2006-NC-105671 .  We thank Andrew Armour, Olivier Buu, Miles Blencowe, Jack Harris, Matt LaHaye and Keith Schwab for useful discussions.

\section*{References}

\bibliographystyle{iopart-num}
\bibliography{references}

%\begin{thebibliography}{2}
%\bibitem{ref:QLimitLinear}A. A. Clerk, Phys. Rev. B 70, 245306 (2004). (cond-mat/0406536)
%\bibitem{ref:QubitsSpectrometers}R. S. Schoelkopf, A. A. Clerk \emph{et. al.,} {}``Qubits as Spectrometers
%of Quantum Noise'' in \emph{Quantum Noise in Mesoscopic Physics,}
%Kluwer, 2003. (cond-mat/0210247).
%\bibitem{ref:DJQPPRL}A. A. Clerk, S.M. Girvin, A. K. Nguyen, and A. D. Stone, Phys. Rev.
%Lett. \textbf{89}, 176804 (2002).
%\end{thebibliography}

\end{document}